# TEACHING QUALITY ASSURANCE AND PROJECT MANGEMENT TO UNDERGRDUATE COMPUTING STUDENTS IN PAKISTAN

─ ABSTRACT ─

**Zaigham Mahmood**
University of Derby, UK
School of Computing, University of Derby, Derby, DE22 1GB, UK
Phone: 0044-1332-540230
E-mail: z.mahmood@derby.ac.uk

**Saqib Saeed**
Bahria University, Islamabad, Pakistan
Shangrilla Road, Sector E-8, Islamabad, Pakistan
Phone: 0092-51-9260002;   Fax: 0092-51-9260885
E-mail: saqib@bci.edu.pk

Software Project Management (SPM) and Software Quality Assurance (SQA) and are key components of undergraduate Computing programmes at educational establishments in Pakistan. Because of the nature of these subjects, there are a number of issues that need to be discussed and resolved so that the teaching becomes more effective, students' learning experience is more enjoyable and their ability to be actively involved in SPM and SQA, after the completion of their studies, becomes further improved. In this paper, we discuss experience of teaching SPM and SQA at one particular institution in Islamabad Pakistan. Using this as a case study, we underline the students' perspective, highlight the inherent issues and suggest ways to improve the delivery of these subjects. Since, the issues are mainly generic, the aim is to provide discussion and recommendations to benefit a wider computing community in academia.

# TEACHING QUALITY ASSURANCE AND PROJECT MANGEMENT TO UNDERGRDUATE COMPUTING STUDENTS IN PAKISTAN

─ FULL PAPER ─

**Zaigham Mahmood**
University of Derby, UK
School of Computing, University of Derby, Derby, DE22 1GB, UK
Phone: 0044-1332-540230
E-mail: z.mahmood@derby.ac.uk

**Saqib Saeed**
Bahria University, Islamabad, Pakistan
Shangrilla Road, Sector E-8, Islamabad, Pakistan
Phone: 0092-51-9260002;   Fax: 0092-51-9260885
E-mail: saqib@bci.edu.pk

## 1. INTRODUCTION

Educational institutions in Pakistan have developed programmes of study in Computing at all levels. The Higher Education Council (HEC) of Pakistan has also been highly instrumental in providing the necessary guidance and assistance. The HEC regard Computing as consisting of a family of disciplines, most notably the areas of Computer Science (CS), Software Engineering (SE) and Information Technology (IT). When studying any of these disciplines, students need to take a number of core modules and some additional options. Two core modules for the CS and SE degrees have been identified as Software Project Management (SPM) and Software Quality Assurance (SQA).

In Pakistan, the software industry is also growing at a steady pace and there is a growing demand for skilled project managers and quality auditors. Although, this demand for expertise in SPM and SQA has resulted in the provision of appropriate programmes of study and the active involvement of the HEC, there is a noticeable lacking in the required skill. A study conducted by Mahmood and Saeed (2008) also noted a distinct lack of expertise in various aspects of Project Management (PM), in particular project planning, risk analysis and risk management. The aim of this paper is to discuss the situation with respect to the delivery of these modules at one particular university in Pakistan, to note issues in the teaching of these modules and suggest ways of improving the delivery of modules so that the teaching becomes more effective and students' learning experience is further enhanced.

In the next section, we discuss the curriculum design of SPM and SQA modules for the degree of Bachelor of Science in SE and CS. In section 3, we highlight the inherent issues with the delivery of the modules. Sections 4 and 5 discuss an experiment that was conducted in 2004-2006 and present our findings with a view to process improvement. Sections 6 and 7 suggest solution strategies and the summary of our findings and recommendations.

## 2. SPM AND SQA CURRICULUM

The structure of the programmes of study at the undergraduate levels and the brief module outlines are defined by the HEC (2004). In the case of the SPM module, the indicative content suggests the teaching of the following:

- PM process and phases
- Stages of PM: initiation, planning, execution, control and monitoring
- Estimation of resources: time, personnel, budget
- Scheduling of activities
- Supporting tools: PERT (Program/Project Evaluation and Review Technique), CPM (Critical Path Method), FPA (Function Point Analysis) and CoCoMo (Constructive Cost Model)
- Management of resources: time, personnel and budget
- Risk analysis
- Standards: project, process and product
- Process improvement

For the SQA module, the indicative content is as follows:

- Quality and quality models

- Software reliability
- Quality assurance: product and process
- Inspections and reviews
- Testing: static, dynamic and automated
- Quality standards
- Quality metrics
- Process improvement
- Certification

Modules content is highly appropriate and in case of SPM, the indicative content appears to follow the well-established Project Management Body of Knowledge (PMBOK) approach (PMI, 2004).

### 3. DELIVERY OF SPM AND SQA MODULES

Our experience relates to the teaching of two particular modules, SQA and SPM, at the final year level of the degrees of BSc in Software Engineering (SE) and BSc in Computer Science (CS), at one particular educational institution in Pakistan. The duration for each of the two modules is 16 weeks and there is a 3-hour traditional face-to-face teaching per week per module. While delivering these modules, an experiment was conducted which started in the fall of 2004 and concluded in the fall of 2006. The process was as follows:

- Observe the way the teaching was carried out and gather student feedback during the first instance of the modules delivery
- Form a subject health group to identify the issues, as a result of the observation and feedback, and design detailed module specifications with recommendations to improve the teaching of modules
- Follow the recommendations for the $2^{nd}$ instance of the delivery of modules, observe the effectiveness and collect further feedback from students
- Make recommendations to further improve and implement recommendations during the $3^{rd}$ instance of modules delivery.

The lessons learnt from one set of delivery of modules were fed into the delivery of subsequent delivery. Thus, there was a continuous process of observation and improvement. In the following two sections, we highlight the issues noted, and the mechanisms implemented, to improve the teaching of these modules.

### 4. ISSUES IN THE TEACHING OF SPM AND SQA

Although, the modules were based on good indicative content, their delivery was less effective because of a number of contributory factors. The main factors are discussed below.

**Inexperience of teaching staff**
Lecturers were usually fresh graduates without or with minimal industrial experience. They knew the theory well but lacked the understanding of the inherent issue with respect to PM and QA (quality assurance). Delivery was often traditional one-way communication so the students, in turn, lacked the practical understanding of the subjects involved. Sometimes, modules were taught by visiting faculty, who often delivered the modules in their own way, not fully concerned with the completion of the syllabus. There was no internal or external moderation of the process of teaching so the teaching strategy was left entirely to the module leaders. Since the industry pays more handsomely to experienced staff, such personnel are less likely to join or move over to academia.

**Overlap of teaching material with other modules**
A considerable overlap was noted between the two modules and other modules, notably, 'Software Engineering-1' and 'Software Engineering-2' which were taught at the same stage. A certain amount of overlap is to be expected, however, when the same member of staff teaches similar modules, overlap can become unacceptably high which is unsatisfactory, especially when the moderation process is non-existing.

**Lack of library resources**
This is a generic problem in all fields of study, in most institutions. Libraries exist but book and journal stock is often inadequate. It was noted that, at the institution where the experiment was being conducted, some subject areas were well stocked, e.g. software engineering, but there were just a few books on PM and QA. Although, books are available in the market and they are not too expensive, there is a limit to how many books can be purchased by students. It should be the responsibility of the education institutions to ensure adequate supply of book and journal stock. This lack of learning resources results in a lack of research and wider reading. Students' knowledge was therefore restricted to what was imparted by the lecturers.

**Incomplete coverage of module content**
This is another generic problem, not restricted to just Computing. Delivery of modules is the responsibility of module lecturers and often there is no check as to how complete, or otherwise, the delivery of the content has been. There is a lack of quality assurance and management at that level which is in the domain of the departments responsible for such programmes of study. For the SPM module, it was noted that size and effort estimation models were not taught and the students had only the superficial knowledge of function point analysis and cost estimation approaches e.g. CoCoMo (Boehm, 2000). The aspects relating to the economic and strategic evaluations of projects, which are highly important for selecting suitable projects for organizations economically and strategically, was also missing. This is partly due to inexperience of the teaching staff, especially the lack of industrial experience. In the case of SQA, topics were covered but they were lacking in depth; process improvement was totally missing and the CMM (capability maturity model) approach was superficially covered. Part of the reason for not covering the module fully, was the high overlap of material with other similar modules, which in turn resulted in a lack of time available to cover the required topics.

**Lack of appropriate practical content**
These modules should include a high practical content, however, the students had no opportunities to engage in a substantial project. Thus, the appreciation of the inherent issues of PM and QA was lacking. Although, students designed test cases, they were too simple to enable the students to appreciate the difficulties of testing complex software. There were not enough opportunities for conducting walkthroughs or static inspections of code. There was virtually no work carried out in a computer laboratory so none of the PM or project-planning tools were used or even demonstrated.

**5. STEPS TAKEN TO IMPROVE THE MODULES DELIVERY**

Problem understood is problem halved. It was not too difficult to enhance the library resource, however, resolution of the issues of inexperienced staff was not possible. Although the concern was raised with the senior management, it is not easy to attract experienced teaching staff, for the reasons mentioned above. Staff training and short internal courses was suggested as a way forward.

Steps were taken to produce detailed module specifications, to ensure that the prescribed content was fully covered and that the students were directed to correct and relevant sources of information via the library resources and the Internet. A subject health group was established who produced workable plans to improve the situation. The mechanisms put in place and some of the steps taken are highlighted below. The following paragraphs provide more details.

**Establishment of a subject health group**
The first major step taken was to set up a subject health group, with a designated subject leader, with the brief that the subject leader would ensure the academic health of the broader subject area, including not just the SPM and SQA modules but other related modules as well e.g. Software Engineering-1 and Software Engineering-2. The subject leader was required to hold three subject meetings per semester. The idea was that: in the first meeting, the module leaders would outline and agree lecture plans for the courses; in the second meeting the progress would be evaluated, which would then be further reviewed in the third meeting, held towards the end of each semester.

**SPM Module design**
Although, the HEC guidelines (HEC, 2004) provide indicative content, the subject health group developed detailed module specifications and agreed to take the PMBOK (project management body of knowledge) approach (PMI, 2004) with the following content:

- Project Initiation (Proposal writing and justification)
- Project Evaluation (Strategic, technical and economic assessment)
- Software size and cost estimation (Function Points, Object Points and CoCoMo)
- Software project planning (Work breakdown structures, Gantt charts, Project Plan development)
- Software risk management (Risk engineering, Risk management Plan development)
- Software quality management (Assessment, monitoring and control)
- Software project resource management (Resource Planning and Monitoring)
- Software project performance tracking and reporting (Gantt chart, Time lines etc., Project reporting)
- Software project configuration management (Change management procedures and Configuration management)
- Software project team management (Leadership, motivation, team building and development)
- Project Closure

**SQA Module design**
Although, the HEC guidelines provide indicative content, the subject health group agreed that two subject areas should provide the focus of this module: software process improvement (SPI) and software testing. Also, the feedback from the alumni students resulted in the following considerations:

- Removal of some of the content overlapping with other software engineering modules
- Provision of a fixed module content, based on certain agreed recommendations
- Increased focus on prevalent national and international standards

- Increased focus on software testing concepts, test case design, testing methodologies and strategies
- More coverage of automated testing and automated testing tools
- Use of PM tools, in particular project-planning software.

The subject health group, who were given the task to look into the detail detailed design of this module, suggested the following topics to be covered:

- Introduction to, and need for SQA
- Quality Assurance and management during the software development life cycle
- SQA as the best practice
- Product, process and project quality
- Reviews and Audits
- Software quality models and standards: McCabe metrics (McCabe, 1994), ISO 9000 family, Assessment Process, ISO 9001, CMM and its variations, Six Sigma and TQM (total quality management)
- CMM and ISO: A comparative analysis
- Measurements and Metrics (for software maintenance, process improvement, software requirement and implementation) and related Issues
- Software Testing (structural, functional and automated testing), best practices, Test cases and test plans.

**Delivery of modules**
The subject health group was also tasked to look into the methods of delivery and assessment. They suggested a more practical approach to teaching including a much higher emphasis on tutorials (for discussions), presentations (by students), practical sessions (for demonstration and use of software tools), guest lectures (for currency of information) and use of case studies (to better understand the difference between theory and practice). These suggestions were followed which enhanced the students' learning experience. It was also proposed that all examination papers and assignment should be internally moderated to ensure that they are set at the right level.

**Establishment of departmental course catalogue**
A departmental course catalogue was established so that complete lecture notes and other teaching materials could be kept for use by new teaching faculty, upon change of module leadership. Thus, good quality ready-made material was available to new faculty at short notice. The idea was that the material would be continuously updated and new subject areas added.

**Use of tutorial exercises and case studies**
Tutorial exercises were developed regarding the business case, project evaluation and estimation techniques components for the SPM module. Case studies were used and discussed. Students were divided into groups and each group was required to present a case study in the class, which followed a positive discussion. Approximately, ten case studies of real situations were provided to/by the students. Arrangements were made for student visits to software houses and other enterprises to gain a first hand experience of industrial practices in the areas of SPM and SQA and to document their findings. Students were required to use case studies, and their observations, to perform gap analyses to analyze the deviations from well-established specific quality standards. This gave students some insight into how things happen in practice. This exposure further strengthened their motivation to learn. As the modules were in the final year of their degree programmes, students were assigned the tasks of developing proposals, project plans and configuration management plans for their final year projects, which allowed students further practical training in the development of such artifacts.

**Student presentations**
Majority of lectures were turned into interactive discussions. Teaching was delivered through a combination of lecture slides, class discussions and student presentations. Students were asked to work in groups to prepare and give presentations on advanced topics of their choice on a related subject matter. This also helped with their communication skills and provided students a feeling of research being carried out in the subject matter, which they highly appreciated.

**Use of laboratories for practical content**
It is important that these modules have a substantial practical content. Initially, Microsoft Project (Microsoft, 2007) was used as a PM tool for the production of Gantts charts, network diagrams, estimation of activities duration and allocation of cost and personnel resources. Students were asked to experiment by changing various parameters to see how this affected other factors (time, budget and human resource). Students' assignment was based on the use of Microsoft Project. This gave students a good insight into the use of PM tools and to apply theoretical knowledge in a practical manner. For the SQA module, a demo of an automated software-testing tool was also arranged so that students could have some appreciation of the automated software testing.

## 6. FURTHER RECOMMENDATIONS

The practical steps taken, such as industrial visits, presentations, software demonstrations, use of case studies and use of PM and QA tools, enhanced students interest and knowledge, as well as their learning experience. As a result, more students chose to do their final year projects in the areas of SPM and SQA.

Although, there was considerable success in our new approach, there is much room for further experimentation and improvement:

- As McDonald suggests (McDonald, 2000) and as evidenced by many organizations, who are more interested in purchasing off the shelf components in their work environments as opposed to developing new software from scratch, SPM curriculum should include a much higher emphasis on software procurement and management of enterprise integration integrating frameworks.
- Courses on PM and QA need to have a much higher practical content and students should be required to manage a small-to-medium sized project in groups so that they can appreciate the inherent issues of PM and QA. However, management of a real project requires a close liaison with industry, as well as additional time which is not always available. It is also useful if students spend a certain amount of time in industry working with practitioners.
- Use of automated tools needs to be encouraged. Numerous software tools exist, developed by various vendors including Microsoft. The Microsoft Office Project (Microsoft, 2007) can be used as a starting point for PM. The course on SQA could include tutorials and use of automated testing tools such as Cantata (IPL, 2008).
- There appears to be a much greater emphasis on out sourcing and getting partner organizations to develop products. Thus PM courses need to include some content on SQA, service level agreements, management and quality issues with respect to software provision by partner organizations.

- There is a greater need to involve the local industry at all levels, as already mentioned above, including curriculum design and work placement for students.

## 7. CONCLUSION

Software Quality Assurance (SQA) and Software Project Management (SPM) are key components of undergraduate Computing programmes at educational institutions in Pakistan. This paper presents the curriculum for SQA and SPM, as defined by the HEC and discusses the way these modules are delivered at one particular institution in Islamabad Pakistan. The paper presents the result of an experiment, which required: 1) observing the way the teaching was carried out during one particular semester, 2) making notes of the issues with respect to modules delivery and students' experience, 3) suggesting ways of improvements for the subsequent delivery of modules, 4) implementing the suggestions during the next semester and making new observations, 5) improving the process further and implementing additional changes for the next instance of modules delivery.

Results of the experiment were encouraging and the experiment can be considered as a success. The aim of our paper is to present our findings and provide discussion and recommendations to benefit, hopefully, a much wider computing community in academia.